\documentclass[]{spie}  
\usepackage[]{graphicx}
\usepackage{amssymb}

\newcommand{\average}[1]{\left \langle {#1} \right \rangle}

\newcommand{\A}{\text{\bf{\large A}}}

\newcommand{\T}{\mathsf{T}} 
\newcommand{\0}{\mathsf{0}} 
\newcommand{\I}{\mathbf{I}} 
\newcommand{\Hkal}{\mathbf{H}_{\infty}} 
\newcommand{\K}{\mathbf{K}_{\infty}} 
\newcommand{\SigmaK}{\mathbf{\Sigma}_\infty} 
\newcommand{\z}{{z}} 

\newcommand{\CovMat}{\mathbf{\Sigma}} 

\newcommand{\N}{\mathbf{N}} 
\newcommand{\D}{\mathbf{G}} 

\newcommand{\dint}{\mathrm{d}} 

\newcommand{\Ad}{\mathcal{A}_\mathsf{tur}} 
\newcommand{\Bd}{\mathcal{B}_\mathsf{tur}} 
\newcommand{\Cd}{\mathcal{C}_\mathsf{tur}} 
\newcommand{\Dd}{\mathcal{D}_\mathsf{tur}} 

\newcommand{\alphavec}{{\mathbf{\alpha}}} 
\newcommand{\betavec}{{\mathbf{\beta}}} 
\newcommand{\rhovec}{\mathbf{\rho}}

\newcommand{\etavec}{{\mathbf{\eta}}} 
\newcommand{\phivec}{{\mathbf{\psi}}} 
 
\newcommand{\varepsilonvec}{{\mathbf{\varepsilon}}} 
\newcommand{\svec}{{\mathbf{s}}} 
\newcommand{\xvec}{{\mathbf{x}}} 
 
\newcommand{\uvec}{{\mathbf{u}}}

\title{Advanced control laws for the new generation of AO systems} 

\author{Carlos M. Correia\supit{a} 
\skiplinehalf
\supit{a}Aix Marseille Univ., CNRS, CNES, LAM, 38 rue F. Joliot-Curie, 13388 Marseille, France; \\
}

\authorinfo{Further author information: (Send correspondence to C.C)\\A.A.A.: E-mail: carlos.correia@lam.fr}

\begin{document} 
  \maketitle 

\begin{abstract}
Geared by the increasing need for enhanced performance, both optical and computational, new dynamic control laws have been researched in recent years for next generation adaptive optics systems on current 10\,m-class and extremely large telescopes up to 40\,m. 
We provide an overview of these developments and point out prospects to making such controllers drive actual systems on-sky. 
\end{abstract}

\section{Introduction}
\label{sec:intro}  
Adaptive optics systems rely on some form of dynamic control to keep up with the evolving atmospheric and telescope disturbances \cite{roddier99, hardy98}. Although historically in closed-loop, open-loop systems have also been deployed on-sky \cite{sivo14, lardiere14}. The developments covered here apply to both configurations.

The need for faster systems with increased performance led to the development of minimum-variance (MV) controllers that can potentially cope with the very large number of degrees of freedom (DoF) in the order of many thousands \cite{correia09, massioni11, massioni14}. The main shortcoming is the computational burden associated and how these solutions can be mapped onto real-time architectures to drive systems at kilohertz frame-rates. 

In this contribution we provide an overview of recent work on optimal and near-optimal solutions for both classical and tomographic systems. We cover ways in which disturbance's temporal evolution can be embedded in the control to improve contrast in extreme AO systems by tackling servo-lag errors causing poorer performance at short separations from the host star. For tomographic systems we reason analogously to provide insight into how spatio-temporal combinations can lead to improved performance.

\begin{figure}
   \begin{center}
   \begin{tabular}{c}
   \includegraphics[height=4cm]{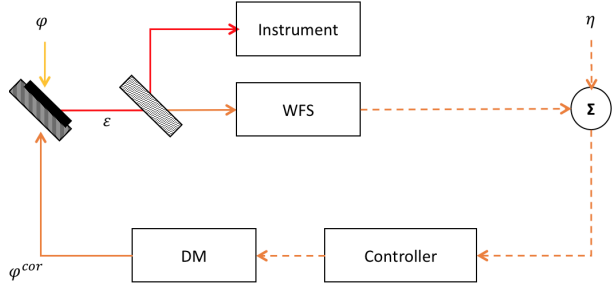}
   \end{tabular}
   \end{center}
   \caption[blockDiag] 
   { \label{fig:blockDiag} 
(Colour on-line) Control-oriented block diagram in closed-loop. The residual error $\varepsilonvec$ is a joint contribution od the phase $\phivec$ and noise $\etavec$. In this paper we focus on the choice of the controller, assuming the rest of the loop to be represented by pure delays.}
   \end{figure}

The use of sub-optimal, \textit{ad-hoc} controllers in AO is commonplace \cite{roddier99, hardy98}. Historically, integral-action controllers were (and are still) the most common choice for a variety of good reasons
\begin{itemize}
    \item AO being a disturbance-rejection dynamical problem \cite{andersonmoore_optimalcontrolLQG05}, driving the measurement error (i.e. the sum of disturbance and propagated noise -- see Fig. \ref{fig:blockDiag}) to zero is a sound approach
    \item ease of manipulation with reduced number of free-parameters \cite{gendron94, dessene98, kulcsar06} -- basically the controller gain
    \item synthetic modelling in open- and closed-loop \cite{ellerbroek05,rigaut98,jolissaint10, correia17} for which closed-forms exist
\end{itemize}
It it worth pointing out that such integral-action control is the first choice to drive the single-conjugate (SCAO) and high-order laser-guide-star assisted tomographic systems on all the three giant telescopes \cite{ramsay13, bouchez14, boyer16}.

For low-order modes however there is a general consensus on the need for more advanced controllers capable of improved performance \cite{correia13, wang14, correia16}. A brief presentation is provided next.
   
\section{Towards Strehl-optimal control}
\label{sec:opticalControl}  

Turning our attention now to Strehl-ratio optimisation (a metric of image quality between 1 for aberration-free diffraction-limited images converging asymptotically to 0 for totally uncorrected
images), it has been shown in Herrman et al (1992) \cite{herrman92} that minimising the residual phase error
variance (MV) is a surrogate for maximising the Strehl-ratio. This makes for a discrete-time minimum residual phase variance (MV) criterion.

To address this class of optimisation problems, we start off by a general discrete-time linear-quadratic (LQ) regulator minimizes the cost function
\begin{equation}
J(\uvec)  = \lim_{M \rightarrow \infty}\frac{1}{M}\sum_{k=0}^{M-1} \left(\xvec^\T \mathbf{Q} \xvec  + \uvec^\T \mathbf{R} \uvec + 2 \xvec^\T \mathbf{S} \uvec \right)_k
\end{equation}
where the weighting triplet $\{\mathbf{Q}, \mathbf{R}, \mathbf{S}\}$ is made apparent and will be specified next by developing an AO-specific quadratic criterion 
subject to the state-space model
\begin{equation}\label{eq:state_space_model}
  \left\{\begin{array}{ccl}
     \xvec_{k+1}  &  = & \Ad \xvec_k  + \mathcal{V}  \mathbf{\nu}_k \\
     \svec_{\alphavec,k}& = & \Cd \xvec_k + \Dd \uvec_{k-d}  + \etavec_k
    \end{array}\right.
\end{equation}
where $\xvec_k$ is the state vector that contains if not the phase directly a linear combination thereof, $\svec_{\alphavec,k}$ are the noisy measurements provided by the WFS in the GS directions $\alphavec$, $\mathbf{\nu}_k$ and $\etavec_k$ are spectrally white, Gaussian-distributed state excitation and measurement noise respectively. In the following we assume that the state drive noise covariance  $\CovMat_{\mathbf{\nu}} = \average{\mathbf{\nu} \mathbf{\nu}^\T}$ and measurement noise covariance $\CovMat_{\mathbf{\mathbf{\eta}}} = \average{\mathbf{\eta} \mathbf{\eta}^\T}$ are known and that $\CovMat_{\mathbf{\mathbf{\eta,\nu}}} = \average{\mathbf{\eta} \mathbf{\nu}^\T} = 0$. Matrices $\Ad, \Bd, \Cd$ will be populated according to the temporal evolution and measurement modes of the AO loop.

The solution to the criterion in Eq. (\ref{eq:Ju}) is known to be a negative state feedback of the form \cite{andersonmoore_optimalcontrolLQG05}
\begin{equation}
    \uvec_k = -\K \xvec_{k}
\end{equation}
where $\K$ is the LQ gain which involves the solution of a Riccati equation \cite{andersonmoore_optimalcontrolLQG05}. The AO problem allows for a simplification leading to an easier to grasp solution.
   
Looking back at the AO optimisation criterion, it takes the form   
\begin{equation}\label{eq:Ju}
J(\uvec) = \lim_{M \rightarrow \infty}\frac{1}{M}\sum_{k=0}^{M-1} \sigma^2_k \nonumber \\
 =  \lim_{M \rightarrow \infty}\frac{1}{M}\sum_{k=0}^{M-1}\Vert
  \phivec_{k+1} (\rhovec,\betavec) - \N \uvec_k  (\rhovec,\betavec) \Vert^2_{W,L_2(\Omega)}
\end{equation}
where $\phivec_{k+1}$ is the average phase disturbance $\in [kT_s:(k+1)T_s[$, $\uvec_k$ is the applied commands over $\in [kT_s:(k+1)T_s[$ producing a correction phase $\N \uvec_k  (\rhovec,\betavec)$ with $\N$ a concatenation of deformable-mirror (DM) influence functions and $\rhovec$ a bi-dimensional spatial index spanning the telescope pupil and $\betavec$ the observation direction of interest. In Eq. (\ref{eq:Ju}), $W$ is a positive-definite weighting matrix that removes the piston contribution over the telescope aperture and $L_2$ the Euclidean norm over the aperture $\Omega$ such that $\sigma^2_k =\Vert
  \varepsilonvec_k\Vert^2_{W,L_2(\Omega)} =\varepsilonvec_k^\T W \varepsilonvec_k$. 
Equation (\ref{eq:Ju}) neglects any WF dynamics during the integration. The error that is neglected here has been dubbed \textit{insurmountable error} due to the use of averaged variables instead of continuous ones \cite{correia10a} and is negligible for the most common systems operating at high frame-rates.

Solution to Eq. (\ref{eq:Ju}) is straightforwardly shown to be a least-squares fit onto the DM influence functions \cite{kulcsar06, correia10a}
\begin{equation}\label{eq:CSI}
\uvec_k  (\rhovec,\betavec) = \mathbf{F}{\phivec}_{k+1}(\rhovec,\betavec) = (\N^\T W\N)^{-1}\N^\T W \phivec_{k+1} (\rhovec,\betavec)
\end{equation}
In passing we note that in order to take into account inter-sample phase and DM dynamics a more involved derivation is necessary \cite{correia10, looze07} leading to adopting instead the general solution in Eq. (\ref{eq:CSI}) as the optimal state feedback gain; this case is not covered here though. Under this somewhat simplified, yet realistic assumption (considering the insurmountable error small and DM bandwidth larger than the AO sampling frequency to be a sound standpoint), the least-squares solution in Eq. (\ref{eq:CSI}) would have been likewise found from the solution of a (albeit degenerate) control Riccati solution \cite{kulcsar12, correia10a}. In this case the weighting triplet $\{\mathbf{Q}, \mathbf{R}, \mathbf{S}\}$ is found in \cite{correia10a} to be $\mathbf{Q} =\mathbf{I}$, $\mathbf{R}= \N^\T W\N$ and $\mathbf{S}=-\N W$ when $\xvec_k=\phivec_{k+1}$. 

In practice, however, $\xvec_{k+1}$ is not known. The separation theorem \cite{andersonmoore_optimalcontrolLQG05} between estimation and control allows writing   
\begin{equation}
\uvec_k  (\rhovec,\betavec)  = \mathbf{F}\widehat{\phivec}_{k+1|k} (\rhovec,\betavec) = (\N^\T W\N)^{-1}\N^\T W \widehat{\phivec}_{k+1|k} (\rhovec,\betavec),
\end{equation}
where $\widehat{\phivec}_{k+1|k} (\rhovec,\betavec) =
\widehat{\phivec}_{k+1|\mathcal{S}_k(\alphavec)}  (\rhovec,\betavec)=
E({\phivec}_{k+1} (\rhovec,\betavec)|\mathcal{S}_k(\rhovec,\alphavec))$ is the conditional expectation of
${\phivec}_{k+1} (\betavec)$ in the science directions  with respect to the sequence of all
measurements from (in general not co-aligned) GS directions available up to $t=k T_s$, \textit{i.e.} $\mathcal{S}_k =
\{\svec_0, \cdots,\svec_k\}(\rhovec,\alphavec)$ 
\cite{andersonmoore_optimalfiltering05}.  

This conditional expectation
is readily available from a Kalman filter
\cite{andersonmoore_optimalcontrolLQG05, correia10a, gilles13}, applied during AO run-time as follows
\begin{equation}\label{eq:steps_KF}
  \left.\begin{array}{ccl}
     \uvec_k & = & \widehat{\xvec}_{k|k-1} \\
     \widehat{\xvec}_{k|k}  &  = & \widehat{\xvec}_{k|k-1} + \Hkal (\svec_k - \Cd \widehat{\xvec}_{k|k-1}) + \Dd \uvec_{k-d} \\
     \widehat{\xvec}_{k+1|k} & = & \Ad \widehat{\xvec}_{k|k}
    \end{array}\right.
\end{equation}
with the asymptotic solutions of the Kalman gain $\Hkal$ found from
\begin{equation}
    \Hkal = \SigmaK\Cd^\T (\Cd \SigmaK\Cd^\T + \Sigma_\eta)^{-1}
\end{equation}
which requires solving for the estimation error covariance matrix $\SigmaK$ from a Riccati equation of the following form
\begin{equation}\label{eq:SigmaKalman}
    \SigmaK = \Ad\SigmaK\Ad^\T +\Sigma_\nu - \Ad\SigmaK\Cd^\T(\Cd\SigmaK\Cd^\T + \Sigma_\eta)^{-1}\Cd\SigmaK\Ad^\T
\end{equation}
This forms the full Linear-Quadradic-Gaussian solution, called LQG controller in the remainder of this paper.

\subsection{Transfer functions}
A common notion is that of transfer function\cite{oppenheim97} which provide the ratio of the output to the input of a system in the Laplace domain (or in 'Z' domain for discrete-time signals). 

Noticing that the residual is a sum of two contributions, namely the residual input disturbance and propagated noise through the loop -- see Fig. \ref{fig:blockDiag}, the residual output variance is computed from 
\begin{equation}\label{eq:res_var_tf}
    \sigma^2_\varepsilonvec = \int_0^{f_s/2} \left|H\left(e^{2i\pi \nu/f_s}\right)\right|^2 \mathbf{W}_\phivec(\nu) \dint \nu + \int_0^{f_s/2} \left|N\left(e^{2i\pi \nu/f_s}\right)\right|^2 \mathbf{W}_\eta(\nu) \dint \nu
\end{equation}
where $f_s=1/T_s$ is the sampling frequency, $\mathbf{W}$ are temporal power-spectral densities (PSDs) of WF and noise. 

The control optimisation problem can be rephrased now as follows. The controller synthesis consists in finding the solution that minimises $\sigma^2_\varepsilonvec$, subject to the Bode's integral theorem
\begin{equation}
    		\int_0^{f_s/2} log \left(H(e^{2i\pi\nu/f_s})\right) \dint \nu = 0.
  	\end{equation}
  	and $\int H(\nu) + N(\nu) \dint \nu = 1$.
  	
  	This is particularly insightful since a control action at a given frequency will have a 'rebound' at another frequency -- the famous waterbed effect. The optimal control solution provides a means to balance the rejection optimally across the temporal spectrum, considering jointly the disturbance and the noise.

Under the assumption of an overall loop transfer taken as a pure delay of $d\times T_s$, $d\in \mathbb{I}$ its transfer function becomes $z^{-d}$, where $z = e^{2i\pi\nu T_s}=e^{2i\pi\nu/f_s}$ is the temporal shift operator,  $d\geq 1$. This yields for the closed-loop rejection functions \cite{roddier99}
\begin{equation}
    H(z) = \frac{1}{1+z^{-d}H_{ctr}(z)}
\end{equation}
for the input disturbance rejection transfer and 
\begin{equation}
    N(z) = z^{-d-1}H_{ctr}(z)
\end{equation}
for the noise transfer function.
with $H_{ctr}(z)$ in the case of the integrator
\begin{equation}
    H_{ctr=int}(z) = \frac{g}{1-z^{-1}}
\end{equation}
where $g$ is the integrator gain. Early attempts to minimise Eq. (\ref{eq:res_var_tf}) by adjusting the controller gain led to the development of the Optimal Modal Gain Integrator \cite{gendron94}, i.e find $g$ that minimises $\sigma^2_\varepsilonvec$.

For the LQG transfer function \cite{correia17}
\begin{equation}
H_{ctr=lqg}(z) = - \left\{\I + \K \Lambda_{e} \left[z^{-1}
        			(\I-\Hkal\Cd) \Bd \right.\right.+ 
                                 \left.\left.\z^{-d}\Hkal\mathcal{D}\right]
    			\right\}^{-1}  
                        \K \Lambda_{e} \Hkal,
	\end{equation}
	where 
  	\begin{equation}
    		\Lambda_{e} = \left(\I - \z^{-1}\Ad + \z^{-1} \Hkal \Cd \Ad \right)^{-1}.
  	\end{equation}
  	
\subsection{Illustrative example}
There are often comments about the overall complexity in the derivation of the optimal control solution in AO (specially compared to the simplicity of the integral controller). During and after this SPIE conference at Austin the same happened. I hope that providing a simple example using single-input single-output (SISO) systems it can make things clearer for anyone willing to delve into optimal control in our community. 

Let's start by the measurement equation
\begin{equation}
    \svec_k = \D \phivec_{k-1} - \N \uvec_{k-2}
\end{equation}
which is the typical 2-step delay closed-loop AO system with $\D$ a linear mapping from wave-front to measurements and $\N$ the mapping from commands to measurements, i.e the interaction matrix. In this SISO example, these are simply scalar values set to one for the sake of simplicity. For reasons that will become apparent momentarily, let the state be
\begin{equation}\label{eq:eg_state}
  \xvec_{k} = \left(\begin{array}{c} \phivec_{k+1}\\\phivec_{k}\\\phivec_{k-1}\end{array}
       \right)
\end{equation}
and state-space matrices
\begin{equation}\label{eq:eg_matrices}
  \mathcal{A} = \left(\begin{array}{ccc} a&b&c\\1&0&0\\0&1&0\end{array}
       \right) 
       \hspace{20pt} 
       \mathcal{C}=\left(\begin{array}{ccc} 0&0&G\end{array}
       \right)
       \hspace{20pt} 
       \mathcal{D}=\begin{array}{c} -GNz^{-2}\end{array}
\end{equation}
where we implicitly chose a model structure up to order 3 ($\{a, b, c\} \neq 0$). The choice of the model order is an important one, since the rejection slope at low-frequencies depends on this value. For the time being, lets pursue under this choice of parameters. Eq. (\ref{eq:CSI}). 

The optimal controls being given by Eq. (\ref{eq:CSI}), we set 
\begin{equation}
    \uvec = \K \widehat{\xvec}_{k|k}= \left(\begin{array}{ccc} 1&0&0\end{array}\right) \widehat{\xvec}_{k|k} = \widehat{\phivec}_{k+1|k}
\end{equation}
which effectively extracts the first element on the state vector, namely $\widehat{\phivec}_{k+1|k}$.

In order to solve for Eq. (\ref{eq:SigmaKalman}) we lack the measurement noise covariance matrix and the state driving noise covariance matrix. 

For the measurement we chose $\Sigma_\eta = \sigma^2_\eta$ a scalar value carrying the noise variance for this single mode measurement. For the state drive noise, $\Sigma_\nu$ must be chosen such that the state output model $x_k$ has the correct variance. The simplest of the cases is $\phivec_{k+1} = a \phivec_k + \nu_k$ leading to $\sigma^2_\nu = (1-a^2)\sigma^2_\phi$. For order 2, i.e. $\phivec_{k+1} = a \phivec_k + b \phivec_{k-1} +\nu_k$ we find $\sigma^2_\nu = (1-a^2-b^2-2a^2b/(1-b))\sigma^2_\phivec$
For higher-order auto-regressive models there is no general closed-form expression and the state drive noise may have to be adjusted numerically. Although model identification is only alluded to here, it is a central topic of research on which the optimal control fully depends \cite{hinnen07}.

We now take a standard atmospheric temporal power spectral density (PSD) \cite{conan95} available for instance from \cite{conan14}. For a $r_0=15$\,cm coherence length atmosphere, with outer scale $L_0=25$\,m and $15\,m/s$ winds-speed, we get 294 $rad^2$. With $a=1.9722$, $b=-0.9724$ and $c=0$ (computed for a model with double pole at 7\,Hz) we find $sigma^2_\nu = 0.0031$. The reader can try out these values to find the correct output variance of the mode as expected. The noise variance is in this example $\sigma^2_\eta=10\,rad^2$.

\begin{figure}
   \begin{center}
   \begin{tabular}{c}
   \includegraphics[height=9cm]{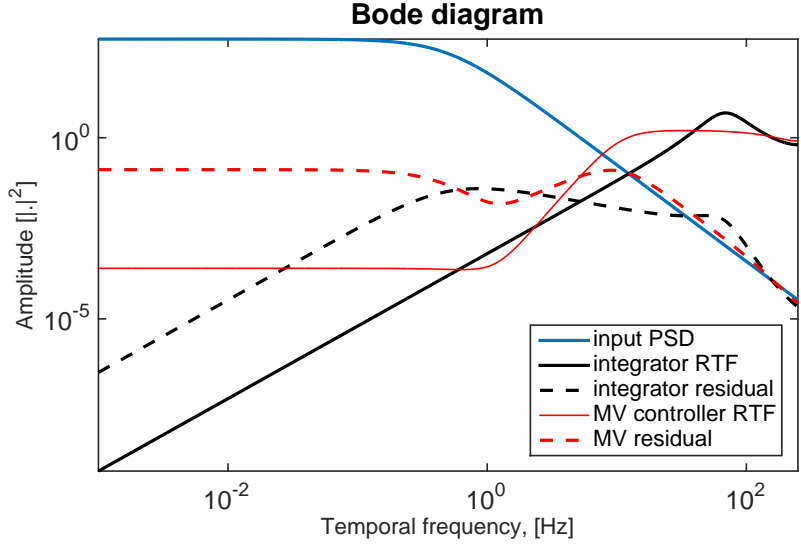}
   \end{tabular}
   \end{center}
   \caption[tfexample] 
   { \label{fig:tf_example} 
(Colour on-line) Example input disturbance and rejection transfer functions for the integrator and LQG controllers.}
   \end{figure}

Figure \ref{fig:tf_example} depicts the input PSD and the rejection transfer functions of the integrator and the optimal controller.

Salient features are:
\begin{itemize}
    \item The optimal controller rejection is proportional to 40dB/dec whereas the integrator is only 20db/dec. This is a tunable parameter that depends on the model order to achieve 20dB/dec $\times$ model-order. 
    \item There is a knee at low frequencies such that the residual PSD is as flat as possible. This depends on the parameters $\{a, b, \cdots\}$ of the model
\end{itemize}


\subsection{Multi-mode case}

From the results above in the mono-mode, scalar case, we can extend the reasoning to a larger set of modes. Initial results were developed in \cite{leroux04, kulcsar06} and later extended by \cite{petit09, correia10, correia15} using a Zernike modal basis\cite{noll76}. Of particular insight is the work by Conan \cite{Conan11} that covers in greater detail than I could possibly do here the application to the multiple-input, multiple-output (MIMO) case. From the observation that the LQG controller provides better rejection for every mode, one can either 
\begin{enumerate}
    \item improve the performance for a given sampling frequency or
    \item relax the sampling frequency, thus integrate longer whilst keeping a performance not worst than the integrator.
\end{enumerate}

Implementations on systems beyond just tip-tilt modes showed promising results on Canary \cite{sivo14} and Raven \cite{lardiere14}.

\subsection{Vibrations}
Vibrations are considered a major source of disturbance affecting almost all observatories around the globe \cite{kulscar12}. Most remarkably, the LQG has been successfully applied to vibration rejection on SPHERE \cite{fusco16} and GPI \cite{poyneer14} using formalism developed in \cite{petit08, meimon10, poyneer10}. 

A continuous-time model is provided in \cite{correia12} which is quite convenient for LGS-based tomographic systems where the low-order modes are corrected separately and potentially at lower frame rates to allow for improved sky-coverage \cite{correia13, wang14}.

In practice the vibration model is concatenated with that in Eq. (\ref{eq:eg_state}) in order to obtain a notch at the vibration central frequencies.

\subsection{From 8\,m to 40\,m class telescopes}
The optimal control features are particularly useful on 40\,m class telescopes, particularly the low-order modes. Since the lowest spatial frequency is $\propto 1/D$, low-order modes will carry lot more disturbance $\propto (D/r_0)^{5/3}$ \cite{Conan11}. This in turn indicates that we need a means to adjust the rejection at very low frequencies along the lines presented in Fig. \ref{fig:tf_example}. Conan et al (2011)\cite{Conan11} provide a full account of the advantages of optimal control on 40\,m class telescopes. 
 
\subsection{The MMSE as a degraded solution of the optimal control}
It has been early recognised that the minimum-mean squared error solution to the (spatial) mode estimation  can be obtained seamlessly from the LQG solution. This is achieved by noting that under a static assumption (i.e. no temporal delays, memoryless model) then $\mathcal{A}=\mathbf{I}$, i.e. $\phivec_{k+1} = \phivec_{k}$ the conditional expectation $\widehat{\xvec}_{k|k-1} = 0$. The estimation error covariance matrix becomes in this case $\SigmaK = \mathbf{\Sigma}_\phivec$ the uncorrected phase covariance matrix leading to
\begin{equation}\label{eq:pol}
    \widehat{\xvec}_{k} = \Hkal (\svec_k + \Dd\uvec_{k-d})
\end{equation}
and as before 
\begin{equation}
    \uvec_k = \mathbf{F} \widehat{\xvec}'_{k|k} = \mathbf{F} \widehat{\phivec}_{k|k} 
\end{equation}
The term $\svec_k + \mathcal{D}\uvec_{k-d}$ represents the so-called pseudo open-loop slopes, i.e. the solution obtained although specified for closed-loop systems makes "as if" it were in open-loop. If instead we were working directly in open loop, then $\Dd = 0$ and the same structure is kept. This is a remarkable feature of the LQG that ultimately stems from the separation principle. 

Since the intrinsic model of the MMSE solution assumes intrinsically $\Ad=\mathbf{I}$ this model is unstable (energy unbound) representative of a random walk. Adaptation of the MMSE to the closed-loop case has been proposed in \cite{gilles05} as a pseudo open-loop control (POLC)  or in \cite{bechet10} -- neither inspired in Eq. (\ref{eq:pol}) --, ensuring on the one hand that the MMSE solution is used (mainly for computational reasons) and  that the solution is stable, although sub-optimal. POLC solutions are the baseline for the AO systems of all giant telescopes, both implementing laser-tomography (as is the case of the ELT and GMT) or multi-conjugation (the case of TMT and the ELT). 
\subsection{Robustness and model identification}
Robustness to changing observing conditions is an item that deserves special care. Identification of models \cite{hinnen07} and adaptation to unstationary conditions can be found in \cite{doelman16}. Recent work is reported in \cite{kooten18}.

\section{Optimal control in large scale AO}
\label{sec:largeScaleAO}  
Although theoretically very appealing, optimal control has several drawbacks for real-time environments:
\begin{itemize}
    \item The filter is recursive meaning that the (k – 1) estimate is needed to update the state. This is by no means fundamentally different from integrator-based update equations coupled or not to pseudo-open-loop measurements calculation;
    \item The gain computation commonly involves the inversion of state-sized matrix equation which by itself is unappealing; 
    \item The state has usually a much higher dimension than the input or output vectors.
\end{itemize}

\subsection{Wide-field AO systems}
Several adaptations to large scale AO systems have been pursued over the years, in particular wide-field AO. 
It has been observed -- and extensively used for modelling purposes -- that the main operations involved in the forward process are, or can reasonably be approximated by, circular convolutions with localized kernels. Here we talk about propagation in weak-turbulence (geometric optics applies, otherwise Fourier optics needed), angular shifts that convert to spatial shifts through easy-to-grasp geometric transformations, temporal shifts which convert correspondingly to spatial shifts on account of the frozen-flow hypothesis and finally, the critical point for modellers, describing the wave-front sensor (WFS) through filtering operations. 

Calculations in the spatial Fourier domain are particularly suitable for the Fourier-transformed bidimensional operators become diagonal. The problem with this approach is that of the spatial invariance of the operators -- it is only an approximation due to the fact that all telescope have finite apertures. This leads to boundary issues (so common in many other fields of physics and engineering) and to circularity concerns. For instance, although the covariance matrices appearing in the formulations admit a quasi-diagonal representation in the spatial Fourier-domain, the matrices are instead Toeplitz with Toeplitz blocks and not circulant with circulant blocks. Rodoplhe Conan points out the exact multiplication of a vector by a Toeplitz matrix in the Fourier-domain \cite{conan14a} a work later extended in \cite{ono18}. 

On the same vein we also note the annular shapes of telescope apertures are not adapted to the use of bi-dimensional Fourier transforms. As argued elsewhere, the Fourier modes are shown to be the eigenmodes of the bi-harmonic equation on a rectangular domain. For annular domains the use of disk-harmonic functions would therefore seem more suited, as they are in turn shown to be the eigenmodes of the bi-harmonic equation on a circle. 

The localised kernels mean in turn that sparse techniques are in principle applicable giving rise to approximate models with a reduced number of non-zero entries \cite{ellerbroek03a}.

In the recent past several research papers on the optimisation of Kalman filters were published. The ideas therein can somehow arbitrarily be categorized as follows, involving both runtime and off-line optimization:
\begin{itemize}
    \item For the main flavours of AO -- single-conjugate (SCAO), ground-layer (GLAO), multi-object (MOAO),  laser-tomography (LTAO) systems --, the need to estimate only a pupil-plane wave-front (possibly averaged over several optimization directions) means that the explicit estimation of layered wave-front profiles can be circumvented leading to considerable computational savings. Resorting to a pixelated description of the wave-fronts (customarily a zonal basis) all the matrices in the spate-space model are sparse but the state transition. Calling for the quasi-Markovianity of the model, reasonable approximations can be found of which bilinear spline phase-shifting operators are a good proxy \cite{correia15}.
    \item Using directly the complex-exponential basis. In this case the modes are temporally and statistically independent (reasonable approximation) which means calculations can be done on a mode-per-mode basis – we're thus under a distributed control framework \cite{massioni15}. It goes without saying that the gain is also computed in such parallel fashion which is chiefly important for off-line but also real-time operations, boiling down to localised convolutions. We note also that, unlike in the previous item, the state-transition matrix is (complex) diagonal. Each entry is a phasor to produce a spatial shift of every mode, mimicking thus wind-blown wave-fronts translating across the pupil \cite{poyneer08, correia17}. 
    \item Using sparse approximations to the discrete-time Riccati equation (DARE) solution and apply the gain by formulating the result as the search for the unknowns x in the Ax = b linear system of equations \cite{correia09}.
    \item Doing a Taylor expansion of the DARE to break the algebraic nature of the Riccati equation, solving for it using a closed-form equation \cite{massioni11}.
    \item Applying the finite-horizon Kalman filter. Instead of using the infinite-horizon steady-state Kalman gain, fixing a sliding, finite-horizon with h steps, and then use at any time step k only the measurements from ($k – 1$) to ($k – h$) to get an estimate. In such case the problem can be cast as a vector-matrix multiplication (VMM). \cite{massioni14}
\end{itemize}

\subsection{High-contrast AO systems}
For high-contrast AO systems, the focus is more on optical performance than beating computational constraints which, for this class of systems, are less of a problem. 

The use of predictive control bears the hope of considerably reducing the servo-lag error. The latter manifest itself as a butterfly-like pattern close to the post-coronagraphic PSF core, much reducing the achievable contrast \cite{guyon17}. The use of the complex exponential basis is particularly adapted on account of the post-coronagraphic PSF and residual error PSD with speckles directly obtained as a scaled version of the PSD \cite{poyneer08, correia17, males18}. Although the models are suitable, identification of the correct parameters and robustness in changing observing conditions are the main limitations to the implementation of predictive control on-sky.  

\section{SUMMARY}
\label{sec:summary}  

Optimal dynamical control has a handful of very welcome advantages, including
\begin{itemize}
    \item Common and non-common path optical aberrations can be directly modelled in the state-space
\item Vibrations can be straightforwardly fed-in 
\item The gains are analytically found, no need to cumbersome dichotomy search routines
\item Dealing with the open or close loop nature of the loops is transparent – pseudo open-loop measurements are obtained at each update step in the latter case when manipulating the Kalman equations
\item The spatio-temporal nature of the AO optimization problem, i.e. the temporal dynamics of the loop is factored in analytically and not following ad-hoc techniques. 
\item The estimation and prediction error variances are found directly from the filter optimization.
\item General-case scenario: optical performance is considerably enhanced 
\end{itemize}
Yet, it also faces several challenges. Of main concern is the computational complexity associated with finding the optimal gain matrices on large scale AO systems, specially tomographic AO using several wave-front sensor probes.

\acknowledgments     
 
The research leading to these results received the support of the A*MIDEX project (no. ANR-11- IDEX-0001- 02) funded by the "Investissements dAvenir" French Government program, managed by the French National Research Agency (ANR).



\end{document}